\documentclass[12pt]{article}
\usepackage{amssymb}
\usepackage{graphicx}
\begin{document}
\title{Chaplygin--like gas and branes in black hole bulks}
\author {Alexander Kamenshchik$^{1,2}$, 
Ugo Moschella $^{3}$ and  Vincent Pasquier$^{4}$.}
\date{}
\maketitle
\hskip-6mm$^{1}$ {\em L. D. Landau Institute for Theoretical Physics,
Russian Academy of Sciences, 2 Kosygina street, 117334, Moscow, Russia.}\\
$^{2}$ {\em Landau Network -- Centro Volta, Villa Olmo, via Cantoni 1,
22100 Como, Italy.}\\
$^{3}$ {\em Dipartimento di Scienze Matematiche Fisiche e Chimiche, Via
Lucini 3, 22100 Como and INFN sez. di Milano, Italy.}\\
$^{4}$ {\em Service de Physique Th\'eorique, C.E. Saclay,
91191 Gif-sur-Yvette, France.}

\begin{abstract}
We explore the possibility to locate a brane in black hole backgrounds. We
study explicitly the cases of BHTZ  and Schwarzschild--anti de Sitter (AdS)
black holes. Our result is that branes cannot be supported by brane tension
alone and it is necessary to introduce other forms of matter on the brane. We
find classes of perfect fluid solutions obeying to peculiar state equations.
For the case of BHTZ bulk geometry the state equation takes exactly the form of
a ``Chaplygin gas'', which is relevant in the brane context. In the
Schwarzschild-AdS case we find new state equations which reduce to
the Chaplygin form when the brane is located near the horizon.
\end{abstract}

\section{Introduction}
In this letter we will study  branes in black hole bulks. We will consider
mainly two examples, namely BHTZ \cite{Banados:1993gq} and anti-de
Sitter--Schwarzschild \cite{Hawking:1983dh} black holes, where we study
classes of branes which are naturally adapted to the black hole geometry at
hand.

The result we will find is that in both cases there is no possible choice of
brane tension so that the geometry under consideration solves the
corresponding  Einstein equations in the bulk spacetime. It is necessary to
go beyond the scheme of Randall and Sundrum \cite{Randall:1999vf} by
introducing  matter on the brane which we will do by making use of perfect
fluids. From a geometrical viewpoint, the reason for this new situation is
that either  the  geometry (in the Schwarschild-AdS case) or the foliations
(in the BHTZ case) of the bulks we are going to study are less symmetric than
in the well-studied cases of \cite{Randall:1999vf,Kaloper:1999sm}. This
excludes the possibility of writing the metric as a genuine warped manifold
\cite{O'Neill,Bertola}. Rather, we are considering a more general setting,
that of ``multi-warped'' manifolds, in which different directions on the brane
are multiplied by different functions of the transverse coordinate.

In the Schwarzschild-AdS case there is an exceptional choice of brane location
where  the possibility exists to consider Einstein equations including only
pure brane tension as in \cite{Randall:1999vf} and no other type of matter.
This particular solution has also appeared recently in \cite{Gomez:2000bu}.
However in our treatment it will appear clearly that this fact is rather a
coincidence because the required equality between energy and opposite of the
pressure is numerical but not functional.

The type of matter found in our investigation are worth commenting. Both in the
BHTZ and Schwarzschild-AdS cases it is possible to identify  state equations
for the fluids on the branes which are universal (in the geometry considered)
in the sense that they are not depending on the choice of brane location. The
matter populating branes located in BHTZ bulks has a state equation that is
particularly simple: the pressure has to be inversely proportional to minus
the (positive) energy density. A fluid satisfying such a state equation is
called ``Chaplygin gas'' . This result is rather interesting since this type
of fluid is precisely of interest in the $d$-brane context
\cite{Jackiw,Ogawa,Goldstone,Hoppe}. In the AdS-Schwarschild case the state
equations we have found are more complicate, but retain the Chaplygin gas form
near the horizon.

\section{BHTZ black holes}

In this section we concentrate on BHTZ black holes. These holes are most easily
constructed  by taking quotients of the three-dimensional anti--de Sitter spacetime
\begin{equation}
AdS_3 = \{x \in {\mathbb R}^4:\ \ {x^0}^2- {x^1}^2 - {x^2}^2 + {x^3}^2 = l^2\},
\end{equation}
w.r.t.  discrete groups of translations along suitable Killing vector fields
\cite{Banados:1993gq}. For this purpose, a two-parameter class
 of coordinate systems for  $AdS_3$ is constructed in the following way \cite{Banados:1993gq}:
\begin{eqnarray}
&& x^0 = l\sqrt{\frac{r^2 - r^2_-}{r^2_+ - r^2_-}} \cosh \left(\frac{r_+\phi}{l} - \frac{r_-t}{l^2}\right),
\nonumber \\
&& x^1  = l\sqrt{\frac{r^2 - r^2_-}{r^2_+ - r^2_-}} \sinh \left(\frac{r_+\phi}{l} - \frac{r_-t}{l^2}\right), \nonumber \\
&& x^2  =  l\sqrt{\frac{r^2 - r^2_+}{r^2_+ - r^2_-}} \cosh \left(\frac{r_+t}{l^2} - \frac{r_-\phi}{l}\right), \nonumber \\
&& x^3  = l\sqrt{\frac{r^2 - r^2_+}{r^2_+ - r^2_-}}
\sinh \left(\frac{r_+t}{l^2} - \frac{r_-\phi}{l}\right),
\label{adscoordinates}
\end{eqnarray}
with $r> r_+ \geq r_-$. In these coordinates the  AdS metric is written as follows:
\begin{equation}
ds^2 = \frac{(r^2-r^2_+)(r^2-r^2_-)}{l^2 r^2} dt^2  -
\frac{l^2 r^2}{(r^2-r^2_+)(r^2-r^2_-)} dr^2 - r^2\left(d\phi - \frac{r_+r_-}{r^2l}dt\right)^2.
\label{adsmetric}
\end{equation}
BHTZ black holes are then obtained by identifying those
points  parametrized by $\phi + 2\pi k$ with integer $k$ \cite{Banados:1993gq}.

Let us introduce a new coordinate $\chi$ as follows:
\begin{equation}
\cosh \frac{2 \chi}{l} = \frac{(r^2 - r^2_+) + (r^2 - r^2_-)}{r^2_+ - r^2_-},
\label{chi}
\end{equation}
which gives
\begin{equation}
d\chi = \frac{lr dr}{\sqrt{r^2 - r^2_+}\sqrt{r^2 - r^2_-}}.
\end{equation}

The use of the  coordinate $\chi$ recasts the metric in a form which is
convenient for a brane ``insertion'':
\begin{equation}
ds^2 = \frac{B^2 \sinh^2 \frac{2\chi}{l}}{A + B \cosh\frac{2\chi}{l}} \frac{dt^2}{l^2}  - \left( {A + B \cosh\frac{2\chi}{l}} \right)
\left(\frac{\sqrt{A^2 - B^2}}{A + B \cosh\frac{2\chi}{l}}\frac{dt}{l}
-d\phi \right)^2 -d\chi^2,
\label{newnmetric}
\end{equation}
where
\begin{equation}
2A = r^2_+ + r^2_-,\;\;\;\;\;2B = r^2_+ - r^2_-.
\end{equation}
It is clear from this expression that the coordinate $\chi$ can play the role
of a transverse space-like coordinate in the bulk as in \cite{Randall:1999vf}.
The difference w.r.t the situation studied in \cite{Randall:1999vf} is that
the metric is now multi-warped, i.e. the remaining terms of $ds^2$ are
multiplied by different functions of $\chi$.

Let us  consider now a brane located at $\chi=\chi_b$. The geometry we are
going to study is made up by two slices of the BHTZ black hole geometry glued
together at the brane with a symmetry (orbifold-type) condition  analogous to
that exploited in  \cite{Randall:1999vf}. This geometry is described by the
following components of the metric tensor (w.r.t. the chosen coordinate
system):
\begin{eqnarray}
&&  g_{tt} =
{B}{l^{-2}}   \cosh \frac{2 u}{l}    - {A}{l^{-2}},\nonumber \\
&& g_{\phi\phi}= -{B}   \cosh \frac{2 u}{l}   - {A},\nonumber \\
&& g_{t\phi} = l^{-1} \sqrt{A^2 - B^2}, \nonumber  \\
&& g_{\chi\chi} = -1.
\label{metric}
\end{eqnarray}
where we have introduced $u = \chi_b + |\chi-\chi_b|$. The contravariant
components are the following:
\begin{eqnarray}
&& g^{tt}  =
\frac{ A + B   \cosh \frac{2 u}{l}}{B^2 \sinh^2 \frac{2 u}{l}}
\ l^2,
 \nonumber\\
&& g^{\phi\phi}  =  \frac{ A - B   \cosh \frac{2 u}{l}}{B^2 \sinh^2 \frac{2 u}{l}},
\nonumber \\
&& g^{t\phi}  = \frac {\sqrt{A^2 - B^2}} {B^2 \sinh^2 \frac{2 u}{l}},
\nonumber \\
&& g^{\chi\chi}  =  -1.
\end{eqnarray}
The metric induced on the brane is obtained as  the restriction of the components given in Eq. (\ref{metric}) (first fundamental form):
\begin{eqnarray}
&&  h_{tt} =
{B}{l^{-2}}   \cosh \frac{2 \chi_b}{l}   - {A}{l^{-2}}, \nonumber \\
&& h_{\phi\phi}= -{B}   \cosh \frac{2 \chi_b}{l}
  - {A},\nonumber \\
&& h_{t\phi} = l^{-1} \sqrt{A^2 - B^2}.
\label{inducedmetric}
\end{eqnarray}
Since the components of the bulk
metric are not continuosly differentiable,
Chistoffel's symbols have finite jumps at $\chi=\chi_b$.
These jumps
give rise to the following Einstein tensor on the brane:
\begin{equation}
G^{(b)}_{ij} = \frac{4}{l^2} \coth \frac{2 \chi_b}{l} \ h_{ij} + \tilde G^{(b)}_{ij},
\end{equation}
where
\begin{equation}
\tilde G^{(b)}_{tt} = -\frac{2B}{l^3 \sinh \frac{2 \chi_b}{l}}, \ \ \ \ \tilde
G^{(b)}_{\phi\phi} = \frac{2B}{l \sinh \frac{2 \chi_b}{l}}, \ \ \ \ \tilde
G^{(b)}_{t\phi}=0.
\end{equation}
Outside the brane the Einstein tensor obviously
coincides with the anti--de Sitter one. We have therefore the following structure:
\begin{equation}
G_{\mu\nu} = G^{AdS}_{\mu\nu} + {\delta}^{i}_{\mu}{\delta}^{j}_{\nu}G^{(b)}_{ij} \delta(\chi-\chi_b).
\end{equation}

Let us first of all consider the non-rotating case. This corresponds to
$2A=2B=r^2_+$ and the metric is static: $g_{t\phi} = 0$ and therefore
$h_{t\phi} = 0$. However, due to the presence of the term $\tilde G^{(b)}$,
the brane tension alone is not enough to solve the corresponding Einstein
equations and it is necessary to introduce matter on the brane. We will
concentrate on perfect fluids (but other types of matter are worth
investigating). The following relations on the brane are thus obtained:
\begin{eqnarray}
&& G^{(b)}_{tt} = (\varepsilon + p) u^2_t - p \ h_{tt}, \nonumber \\
&& G^{(b)}_{\phi\phi} = (\varepsilon + p) u^2_\phi - p \ h_{\phi\phi},
\nonumber\\
&& G^{(b)}_{t\phi} = (\varepsilon + p) u_t u_\phi = 0.
\end{eqnarray}
The third equation calls for a static fluid, corresponding to the condition
$u_\phi=0$. The first two equations in the previous array  become simply
$G^{(b)}_{tt} = \varepsilon \ h_{tt}$  and $G^{(b)}_{\phi\phi} =  - p \
h_{\phi\phi}$, which are solved by
\begin{eqnarray}
&& \varepsilon = + \frac{2}{l} {\tanh \frac{ \chi_b}{l}}   \nonumber \\
&& p = -\frac{2}{l} {\coth \frac{ \chi_b}{l}}
\label{state}
\end{eqnarray}
It is possible to write an equation of state which includes all
the previous solutions:
\begin{equation}
p \, \varepsilon = -\frac{4}{l^2}. \label{state1}\end{equation} This state
equation is universal in the sense that it is the only one which has the same
form for any brane location. It corresponds formally to the so-called Chaplygin
gas which has recently raised a certain interest \cite{Jackiw,Ogawa}. The
pressure of this fluid is negative and inversely proportional to the energy
density. The interesting fact is that this type of fluid can be obtained also
from the Nambu-Goto action for $d$-branes moving in a $(d+2)$-dimensional
spacetime in the light-cone parametrization \cite{Goldstone,Hoppe}. For a
string embedded in a three-dimensional flat spacetime the argument is so simple
that we may reproduce it:  the relevant Hamiltonian is $$H= \frac12\int
\left[P^2+(\partial_{\sigma} x)^2 \right] d\sigma, $$ where $x$ is the only
transversal coordinate. Let us introduce euristically the density \cite{Hoppe}
$\varepsilon(x)=(\partial_{\sigma} x )^{-1}$ and the velocity $v=P$. One of
the equations of motion is then current conservation:
$$\partial_t\varepsilon+\partial_x(\varepsilon v)=0.$$ The other gives $$\dot
P=\partial_{\sigma}^2x=\partial_{\sigma}x\partial_x(1/\varepsilon)$$ that is
$$\varepsilon\dot
P=\varepsilon(\partial_t+v\partial_x)v=\partial_x(1/\varepsilon)$$ which is
Euler equation for a fluid such that $p=-1/{\varepsilon}$.

Let us pass now to a short discussion of the rotating case. Einstein equations
now give the following relations on the brane:
\begin{eqnarray}
&& G^{(b)}_{tt} = (\varepsilon + p) u^2_t - p \ h_{tt}, \nonumber \\
&& G^{(b)}_{\phi\phi} = (\varepsilon + p) u^2_\phi - p \ h_{\phi\phi},
\nonumber\\
&& G^{(b)}_{t\phi} = (\varepsilon + p) u_t u_\phi - p \ h_{t\phi}.
\end{eqnarray}
This system should be  supplemented by the normalization condition
\begin{equation}
h^{tt} u_t^2 + 2h^{t\phi} u_t u_\phi + h^{\phi\phi} u_\phi^2 = 1.
\end{equation}
After some algebraic computation the following solution is obtained:
\begin{equation}
u_t^2 = \frac{r^2_+}{l^2} \sinh^2 \frac{\chi_b}{l},\,\,\,\,
u_\phi^2 = {r^2_- \sinh^2 \frac{\chi_b}{l}}.
\end{equation}
Now we can proceed as before and solve for the fluid energy density  and
pressure. The solution, and consequently the equation of state, are exactly
the same as in the static case, see Eqs. (\ref{state},\ref{state1}). The
corresponding full energy-momentum tensor is now  not diagonal.

The coincidence of the energy and pressure obtained in the rotating and
nonrotating cases deserves a further explanation. A closer look to Eq.
(\ref{chi}) reveals that the brane does not really depends on the choice of
$r_+$ and $r_-$ and consequently of the corresponding Killing vectors. Indeed,
from the viewpoint of the space in which the anti-de Sitter universe is
embedded, the brane can be decribed by the following equations:
\begin{equation}
\left\{
\begin{array}{l}
{x^0}^2 - {x^1}^2 + {x^2}^2 - {x^3}^2 = l^2 \cosh \frac {2 \chi_b}{l} \\
{x^0}^2 - {x^1}^2 - {x^2}^2 + {x^3}^2 = l^2
\end{array}
\right.
\label{braneeq}
\end{equation}
 This means that we could have studied such a brane in
the anti-de Sitter bulk {\it before} making identifications along the Killing
vectors. The simplest choice amounts to studying a bulk spacetime with interval
\begin{equation}
ds^2 = \sinh^2 u \ dt^2 - \cosh^2 u \ d\phi^2 - d\chi^2
\label{simple}
\end{equation}
which describes the geometry of two regions of the anti de Sitter manifold
glued together at a two-dimensional brane having the topology of a plane. The
BHTZ procedure then changes both the bulk and the brane topology but has no
consequence on the energy-momentum tensor of the matter living on the brane.

It is interesting to note that the brane is obtained in Eq. (\ref{braneeq}) as
the intersection of two AdS spacetimes embedded in $\mathbb R^4$. Of course
this is peculiar to two-dimensional branes. The present study could in
principle be extended to other dimensionalities \cite{multi}, but in this case
the fluid would not be isotropic.

\section{Schwarzschild-AdS black holes}

\begin{figure}[h]
\begin{center}
\includegraphics[height=6cm]{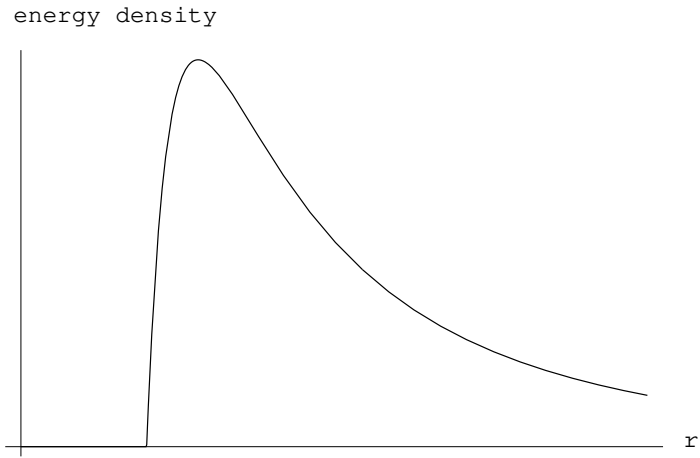}
\caption{Energy density $\varepsilon$ as a function of the brane location for finite positive values of $l$ and $M$. The energy density $\varepsilon$ vanishes when the brane is located at the horizon; $\varepsilon$ has a maximum at $r_b^2=4M$. } \label{fig1}
\end{center}
\end{figure}

We now examine the situation for the Schwarzschild-AdS metric
\cite{Hawking:1983dh}, as recently studied also in \cite{Gomez:2000bu}. We do
this in the usual case in which the bulk spacetime is five-dimensional and the
brane four-dimensional. The  Schwarzschild-AdS metric \cite{Hawking:1983dh} is
given by
\begin{equation}
ds^2 = \left(1 + \frac{r^2}{l^2} - \frac{2M}{r^2}\right)dt^2 -
\frac{1}{1 + \frac{r^2}{l^2} - \frac{2M}{r^2}} \ dr^2 - r^2 d^2\Omega^{(3)},
\label{adssch}
\end{equation}
where $d^2\Omega^{(3)}$ is the metric of a unit three-dimensional spherical
surface. As before we may introduce a radial coordinate $\chi$ by
\begin{equation}
\cosh\frac{2\chi}{l} = \frac{2r^2 + l^2}{\sqrt{8Ml^2 + {l^4}}},
\label{chisch}
\end{equation}
and insert a ``brane'' at $\chi= \chi_b(r_b)$ by requiring the same symmetry
properties of the metric w.r.t. $\chi_b$ as in the previous section. The brane
is now substantially a copy of a Einstein static universe. The Einstein tensor
includes a term proportional to  $\delta(\chi-\chi_b)$, which should be
matched by matter on the brane so as to compensate it. Again, a perfect fluid
with the following energy density and pressure does the job:
\begin{eqnarray}
&& \varepsilon = \frac{6}{r_b}\sqrt{1 + \frac{r_b^2}{l^2} - \frac{2M}{r_b^2}},
\label{energy} \\
&& p = -\frac{4\left(1 +\frac{3r_b^2}{2l^2} -\frac{M}{r_b^2}\right)}
{r_b\sqrt{1 + \frac{r_b^2}{l^2} - \frac{2M}{r_b^2}}}.
\label{pressure2}
\end{eqnarray}
The state equation now depends on the black hole mass $M$:
\begin{equation}\label{state2}
p= -\frac{\varepsilon}{3} - \frac{24}{\varepsilon l^2}
-\frac{12}{\varepsilon r^2_b(\varepsilon)}
\end{equation}
where $r^2_b(\varepsilon)$ is obtained by solving Eq. (\ref{energy}). Two
regimes are possible: if $0<\varepsilon < 6/l$ there is only a solution, given
by
\begin{eqnarray}\label{repsilon1}
&& r^2_b(\varepsilon) = -\frac{18 l^2}{36 - \varepsilon^2 l^2} +
\sqrt{\left(\frac{18 l^2}{36 - \varepsilon^2 l^2}\right)^2 +
\frac{72 M l^2}{36 - \varepsilon^2 l^2}}
\end{eqnarray}
When $\varepsilon > 6/l$ the following two solutions are possible:
\begin{eqnarray}\label{repsilon2}
&& r^2_b(\varepsilon) = \frac{18 l^2}{\varepsilon^2 l^2 - 36} \pm
\sqrt{\left(\frac{18 l^2}{\varepsilon^2 l^2 -36}\right)^2 -
\frac{72 M l^2}{\varepsilon^2 l^2-36}}
\end{eqnarray}
In this regime there is a special value $r_b^2=4M$ where to locate the
brane. Here, the energy density $\varepsilon$ has a maximum value
\begin{equation}
\varepsilon_{\rm max}  = \frac{6}{l}\sqrt{1+\frac{l^2}{8M}},
\end{equation}
and the two solutions coalesce; beyond this value of the energy density there
is no solution (see figure \ref{fig1}).

Since in this very case one also has that $\varepsilon = -p$ numerically, the
fluid can be alternatively interpreted as tension of the brane as in the
original work by Randall and Sundrum \cite{Randall:1999vf}. This particular
situation has been recently considered by \cite{Gomez:2000bu}. However, in our
scheme it appears clearly that this case is rather special and in any other
location one is obliged to add other matter contributions on the brane.

It is worthwhile to remark that the state equation (\ref{state2}) reduces  to
the Chaplygin equation when the brane is located near the horizon. For
$r_b^2>4M$ there exists also the possibility to think of two distinguished
types of matter on the brane, namely brane tension and a standard fluid having
both energy density and pressure positive.

When $l$ tends to infinity we are describing a brane in a Schwarzschild bulk.
Eqs. (\ref{energy}), (\ref{pressure2}), (\ref{state2}) and (\ref{repsilon2}) can be easily reduced to this case; the situation is substantially similar to that we have described above in the Schwarschild-AdS case and we do not reproduce the results.

On the other side, when $M$ is set to zero the metric (\ref{adssch}) reduces to the metric of an
empty AdS spacetime and we are back to the Randall-Sundrum setting.
However, by putting a brane at some value $r = r_b$  of the chosen coordinates
we are exploring the different case in which the brane has the
four-geometry of an Einstein static universe. The constant $r_b$ has therefore
also the meaning of the radius of this universe. Also in this case we need to
add matter beyond brane tension and this because of the brane geometry. Energy
and pressure of the matter populating this universe have a comparatively
simple form:
\begin{eqnarray}
&& \varepsilon = \frac{6}{r_b}\sqrt{1 + \frac{r_b^2}{l^2}}, \label{energy1} \\
&& p = -\frac{4\left(1 +\frac{3r_b^2}{2l^2}\right)}
{r_b\sqrt{1 + \frac{r_b^2}{l^2}}}.
\end{eqnarray}
The state equation is now as follows:
\begin{equation}
p = -\frac{2\varepsilon}{3}  - \frac{12}{\varepsilon l^2}.
\label{Chapl}
\end{equation}
When the radius $r_b$ of the brane tends to infinity we are finally back to
the original Randall-Sundrum case \cite{Randall:1999vf}. Indeed in this
limit we have that numerically
\begin{equation}
\varepsilon  = \frac{6}{l}= - p.  \label{energy12}
\end{equation}
Eq. (\ref{energy12}) recovers the relation betweeen the brane tension and the
bulk cosmological term given in \cite{Randall:1999vf} (up to notations).

Finally, the  state equation(\ref{Chapl}) can be easily generalized to the
$(n+2)$-dimensional anti de Sitter bulk spacetime as follows:
\begin{equation}
p = -\frac{\varepsilon(n-1)}{n} - \frac{4n}{\varepsilon l^2}.
\label{general}
\end{equation}

\section{Concluding remarks}

We have shown that  the study of branes in black hole bulks requires
the use of multi-warped manifolds and there is the necessity of
matter on the branes to set up acceptable Einstein equations.

The matter we find is rather interesting. It is a ``Chaplygin gas'' in BHTZ
situation and a more complicate type of gas in the Schwarschild-AdS case, which
however retains the ``Chaplygin'' form near the horizon.   In both cases the
matter we find is of ``cosmological'' type: a (positive) cosmological can be
thought as a fluid having (constant) positive energy density and negative
pressure. This type of matter may also be of interest in the context of
``quintessential'' expansion models and, from an observative viewpoint, it may
have a connection with the by now accepted existence of a positive
acceleration of the universe expansion \cite{Bachall,Sahni:1999gb}.

It would be interesting to explore if there exist an interpretation for the
equation of state we have found in the Schwarschild-AdS case from the
viewpoint of a theory of extended objects.

\section*{Acknowledgements}

We are grateful to V. Gorini for useful discussions.
A.K. is grateful to the CARIPLO Science Foundation for the financial
support. His work was also partially supported by RFBR via grant No
99-02-18409.

\end{document}